%% file: GW_neutrino_GW151226.tex
\documentclass[aps,prd,reprint,superscriptaddress]{revtex4-1}
\usepackage{graphicx}
\usepackage{enumerate}

\usepackage{subfigure}
\usepackage{amsmath}
\usepackage{amssymb}
\usepackage{amsfonts}
\RequirePackage{lineno}
\usepackage{color}
\usepackage[normalem]{ulem}
\usepackage{soul}
\usepackage{lineno}
\usepackage[table]{xcolor}
\usepackage{hyperref}
\hypersetup{
    pdfnewwindow=true,      
    colorlinks=true,        
    linkcolor=blue,         
    citecolor=black,        
    filecolor=blue,         
    urlcolor=blue           
}

\def\be{\begin{equation}}
\def\ee{\end{equation}}

\providecommand{\aap}[0]{Astron. Astrophys. }

\providecommand{\astropart}[0]{Astropart. Phys. }
\providecommand{\apj}[0]{Astrophys. J.~}
\providecommand{\apjl}[0]{Astrophys. J. Lett. }
\providecommand{\apjs}[0]{Astrophys. J. Supp. Ser. }

\providecommand{\epjc}[0]{Eur. Phys. J. C } 

\providecommand{\ji}[0]{J. Instrum. } 
\providecommand{\jpcs}[0]{J. Phys. Conf. Ser. }
\providecommand{\mnras}[0]{Mon. Not. Roy. Astron. Soc. }

\providecommand{\nimA}[0]{Nucl. Instrum. Meth. A } 

\providecommand{\prd}{Phys. Rev. D. }
\providecommand{\prl}[0]{Phys. Rev. Lett. }
\providecommand{\rpp}[0]{Rep. Prog. Phys. } 

\begin{document}


\title{Search for High-energy Neutrinos from Gravitational Wave Event
GW151226 and Candidate LVT151012 with ANTARES and IceCube}

  \let\mymaketitle\maketitle
  \let\myauthor\author
  \let\myaffiliation\affiliation
  \author{ANTARES Collaboration}
  \author{IceCube Collaboration}
  \author{LIGO Scientific Collaboration}
  \author{Virgo Collaboration}
  \thanks{Full author list given at the end of the article.}



\begin{abstract}
The Advanced LIGO observatories detected gravitational waves from two binary black hole mergers during their first observation run (O1). We present a high-energy neutrino follow-up search for the second gravitational wave event, GW151226, as well as for gravitational wave candidate LVT151012. We find 2 and 4 neutrino candidates detected by IceCube, and 1 and 0 detected by \textsc{Antares}, within $\pm500$\,s around the respective gravitational wave signals, consistent with the expected background rate. None of these neutrino candidates are found to be directionally coincident with GW151226 or LVT151012. We use non-detection to constrain isotropic-equivalent high-energy neutrino emission from GW151226 adopting the GW event's 3D localization, to less than $2\times 10^{51}-2\times10^{54}$\,erg.
\end{abstract}

\pacs{Valid PACS appear here}
\maketitle


\section{Introduction}

Gravitational wave (GW) astronomy began with the observation of a binary black hole (BBH) merger by Advanced LIGO on September 14\textsuperscript{th}, 2015 \cite{G184098}. Following this first discovery, LIGO recorded an additional BBH merger, GW151226 \cite{2016PhRvL.116x1103A}. Another possible signal, named LVT151012, has also been identified with $87\%$ probability that it was of astrophysical origin \cite{2016arXiv160604856T}. These events provide information on the formation mechanism, environment and rate of BBH mergers. They also enable sensitive tests of gravity in the strong field regime \cite{2016arXiv160604856T}.

The GW signals were followed up by a broad multimessenger observation campaign, covering the full electromagnetic spectrum \cite{2016ApJ...826L..13A} as well as neutrinos \cite{2016PhRvD..93l2010A,2016arXiv160807378T,2016ApJ...830L..11A}. Data from the Gamma-ray Burst Monitor on the Fermi satellite \cite{2016ApJ...826L...6C} indicate a signal that could be associated with the first merger observed, GW150914, although this signal is in tension with non-detection by INTEGRAL \cite{2016ApJ...820L..36S}. BBH mergers may produce electromagnetic or neutrino emission if a sufficient amount of circumbinary matter is available for accretion. Most BBH systems likely lack such an environment; however, some binaries residing in active galactic nuclei \cite{2016arXiv160203831B,2017MNRAS.464..946S}, or those with gas remaining from their stellar progenitors \cite{2016ApJ...821L..18P,2016ApJ...819L..21L}, may produce a detectable counterpart \cite{2016ApJ...822L...9M,2016ApJ...827L..16L}.

Accreting black holes can drive relativistic outflows \cite{2006RPPh...69.2259M}. Dissipation within outflows with a hadronic component can produce non-thermal, high-energy neutrinos \cite{1997PhRvL..78.2292W,2010MNRAS.407.1033B}.

High-energy neutrinos of astrophysical origin have recently been discovered by the IceCube detector \cite{2014PhRvL.113j1101A,2015PhRvL.115h1102A,2015PhRvD..91b2001A,2015ApJ...809...98A}, however, the source of these neutrinos is currently unknown.

In this paper we report the results of high-energy neutrino follow-up searches of GW event GW151226 and GW candidate LVT151012 using the IceCube Neutrino Observatory, a cubic-kilometer facility at the South Pole \cite{2006APh....26..155I,2009NIMPA.601..294A,2010NIMPA.618..139A}, and the \textsc{Antares} neutrino telescope in the Mediterranean sea \cite{ANTARES,2007NIMPA.570..107A,2005NIMPA.555..132A}. We briefly discuss the detectors and search procedure in Section \ref{sec:search}, and present the results in Section \ref{sec:results}. We summarize our results and conclude in Section \ref{sec:conclusion}.

\section{Analysis}
\label{sec:search}

On December 26, 2015 at 03:38:53 UTC, the Advanced LIGO detectors observed the coalescence of two black holes, an event named GW151226, with estimated masses of $14.2^{+8.3}_{-3.7}$M$_{\odot}$ and $7.5^{+2.3}_{-2.3}$\,M$_{\odot}$, at a luminosity distance of $440^{+180}_{-190}$\,Mpc, corresponding to a redshift of $0.09^{+0.03}_{-0.04}$ \cite{2016arXiv160604856T}. Subsequently, the significance of the event was established to be greater than $5\,\sigma$ by off-line analyses. The source of the GW was confined to within 850\,deg$^2$ of the sky at 90\% credible level (hereafter skymap) \cite{2016arXiv160604856T}.

Beyond GW151226 (and the first observed GW event GW150914 \cite{2016PhRvL.116f1102A}), LIGO also detected a GW event candidate, LVT151012, on October 12, 2015 at 09:54:43 UTC \cite{2016arXiv160604856T}. While this candidate was not sufficiently significant to claim discovery, it is probably of astrophysical origin. If LVT151012 is indeed a GW signal, it is consistent with a BBH merger at luminosity distance 1000$^{+500}_{-500}$\,Mpc, or redshift of 0.20$^{+0.09}_{-0.09}$, with black hole masses of $23^{+18}_{-6}$M$_\odot$ and $13^{+4}_{-5}$M$_\odot$. The source direction was confined to a 1600\,deg$^2$ skymap \cite{2016arXiv160604856T}. Since this event candidate is probably astrophysical, we include it in this analysis.

We searched for neutrinos coincident with GW151226 and LVT151012 using a time window of $\pm 500$\,s around the GW transients. This is our standard search window adopted for joint GW-neutrino searches \cite{2011APh....35....1B}. Within the $\pm 500$\,s, we do not further weigh the temporal difference between GWs and neutrinos. This time difference, nevertheless, may be indicative of the underlying emission mechanism \cite{2003PhRvD..68h3001R,2012PhRvD..86h3007B,2014PhRvD..90j1301B}.

For IceCube, we adopted the detector's online event stream, which is used in IceCube's online analyses \cite{1742-6596-718-6-062029,2016arXiv161206028I}. This event selection was adopted to ensure compatibility with low-latency GW+neutrino searches. The online event stream uses an event selection similar to that of point source searches \cite{2014ApJ...796..109A}, but is optimized for near-real-time analysis at the South Pole.  This event selection consists primarily of cosmic-ray-induced background events, with an expectation of 2.2 events in the northern sky (atmospheric neutrinos) and 2.2 events in the southern sky (high-energy atmospheric muons) per 1000 seconds. In the search window of $\pm 500$\,s centered on the GW alert times, 2 and 4 neutrino candidates were found by IceCube in correspondence of GW151226 and LVT151012, respectively. This result is consistent with the expected background. The properties of these events are listed in Table~\ref{table:neutrinos}. The listed muon energies are reconstructed assuming a single muon is producing the event. The sky location of the neutrino candidates are shown in Fig. \ref{fig:skymap}. The significantly greater reconstructed energy for the neutrino candidates on the southern hemisphere is consistent with our expectations due to the different selection criteria on the two hemispheres, allowed by the Earth's filtering effect of atmospheric muons.

We performed an additional search for high-energy starting events detected by IceCube (that is, events with tracks starting within the detector). A significant fraction of high-energy starting events are likely of astrophysical origin given the low background rate at the considered high energies. The corresponding IceCube event selection is described in \cite{2014PhRvL.113j1101A}. No high-energy starting events were found in coincidence with GW151226 or LVT151012.

The IceCube detector is also sensitive to outbursts of MeV neutrinos via a sudden increase in the photomultiplier counting rates. Galactic core-collapse supernovae, e.g., will be detected with high significance \cite{2011A&A...535A.109A}. This global counting rate is monitored continuously, the influence of cosmic-ray muons is removed and low-level triggers are formed when deviations from the nominal rate exceed pre-defined levels. An IceCube MeV neutrino trigger was issued on October 12\textsuperscript{th}, 2015, 09:56:36 UTC. The probability of a trigger with the recorded excess counting rate to occur during the $\pm500$\,s time-window around the GW candidate is 12\%. This is not sufficiently significant to require further consideration. To account for the possible time delay of $\sim$\,MeV neutrinos traveling from the reconstructed distances of GW151226 and LVT151012, we also considered an extended time window of $\pm1$\,h. For LVT151012, the same trigger remained the most significant even within this extended window. For GW151226, the trigger with the highest excess counting rate within $\pm1$\,h was recorded +51\,min after the GW event. Events with at least the measured excess counting rate occur at a rate of $\sim0.3$\,h$^{-1}$, therefore we do not consider it to be of astrophysical origin.

\begin{table*}
\begin{tabular}{c|c|c|c|c|c|c|c}
  \hline
  Event     & $\#$ & Detector & $\Delta T$ [s] & RA [h]   & Dec [$^\circ$] & $\sigma_{\mu}^{\rm rec}$ [$^\circ$] & E$_{\mu}^{\rm rec}$ [TeV] \\ \hline
  GW151226  & 1    & \textsc{Antares} & $-387.3$       &  16.7   & $-28.0$        & 0.7                                 &   9                \\
  GW151226  & 2    & IceCube          &  $-290.9$      &  21.7   & $-15.1$        & 0.1                                 & 158              \\
  GW151226  & 3    & IceCube          &   $-22.5$      &   5.9   &   14.9         & 0.7                                 &   6.3          \\ \hline
  LVT151012 & 1    & IceCube          &  $-423.3$      &  24.0   &   28.7         & 3.5                                 &   0.38           \\
  LVT151012 & 2    & IceCube          &  $-410.0$      &   0.5   &   32.0         & 1.1                                 &   0.45            \\
  LVT151012 & 3    & IceCube          &   $-89.8$      &   7.7   & $-14.0$        & 0.6                                 &  13.7              \\
  LVT151012 & 4    & IceCube          &   $147.0$      &   0.6   &   12.3         & 0.3                                 &   0.35             \\
  \hline
\end{tabular}
  \caption{Parameters for neutrino candidates detected by IceCube within $\pm 500$\,s around GW151226 and LVT151012. $\Delta T$ is the time of arrival of the neutrino candidates relative to that of the GW event. E$_{\mu}^{\rm rec}$ is the reconstructed muon energy. $\sigma_{\mu}^{\rm rec}$ is the (50\% for IceCube, 1$\sigma$ for \textsc{Antares}) angular uncertainty of the reconstructed track direction \cite{2014PhRvD..89j2004A,2014ApJ...786L...5A}. 
  }
  \label{table:neutrinos}
\end{table*}

We searched for coincident neutrinos within \textsc{Antares} data by selecting up-going events. The search was performed with the most recent official offline data set, produced incorporating dedicated calibrations, in terms of positioning \cite{2012JInst...7T8002A}, timing \cite{2011APh....34..539A}  and efficiency \cite{2007NIMPA.570..107A}. This sample is dominated by background events from mis-reconstructed down-going atmospheric muons. It was optimized for each GW event individually so that one event that passes the search criteria and is located within the 90\% GW probability contour would lead to a detection with a significance level of $3\sigma$. For GW151226, a total of $1.4\times10^{-2}$ atmospheric neutrino candidates are expected in the field of view within $\pm500$\,s, while the number of misreconstructed down-going muons amounts to $8\times10^{-2}$ events in the same time window. We found one event that is temporally coincident with GW151226, located outside the 90\% GW probability contour. The Poissonian probability of detecting at least one such background event when $9.4\times10^{-2}$ are expected is $\sim9\%$. Thus, this detection is consistent with the expected background muon rate and we conclude that this event is likely a misreconstructed down-going muon. The properties of this event are listed in Table \ref{table:neutrinos}. In particular, the estimated deposited energy \cite{2013EPJC...73.2606A} is 9\,TeV, in agreement with what is expected from a misreconstructed down-going muon. The sky location of the event is shown in Fig. \ref{fig:skymap}.

For LVT151210, the atmospheric neutrino candidate rate expected from the southern sky within $\pm500$\,s is equal to $1.8\times10^{-2}$ while the number of misreconstructed down-going muons amounts to $4\times10^{-2}$. These are somewhat different from the values obtained for GW151226 as the sensitivity of \textsc{Antares} varies with time. No neutrino candidates temporally coincident with LVT151012 were found with \textsc{Antares}.

\section{Results}
\label{sec:results}

\subsection{Constraints on neutrino emission}

\begin{figure}
\begin{center}
\resizebox{0.49\textwidth}{!}{\includegraphics{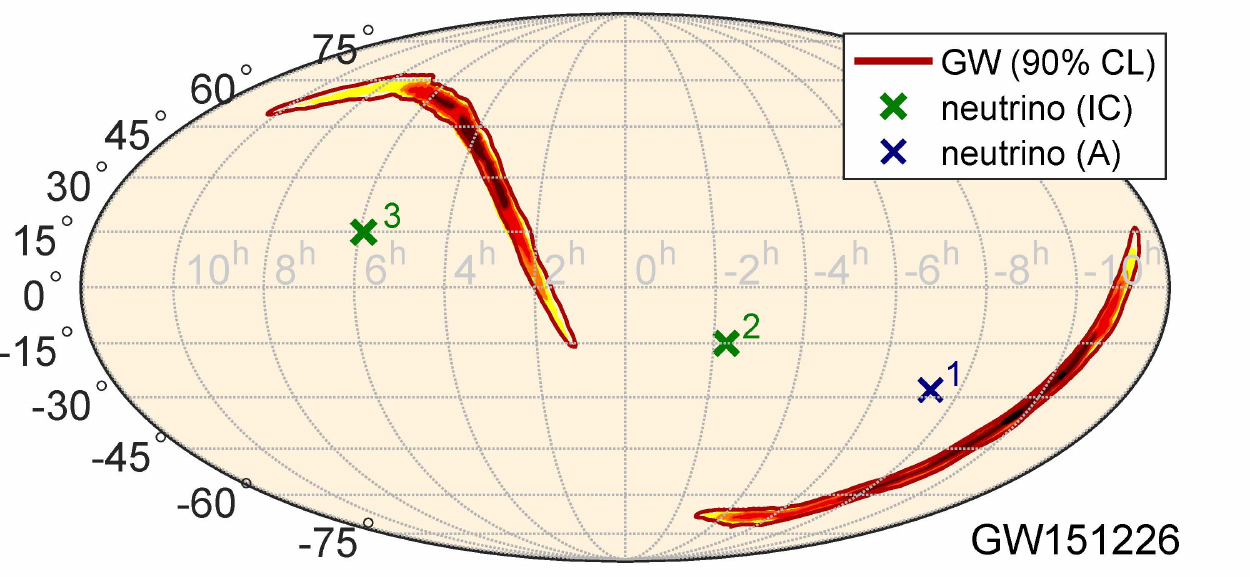}}
\resizebox{0.49\textwidth}{!}{\includegraphics{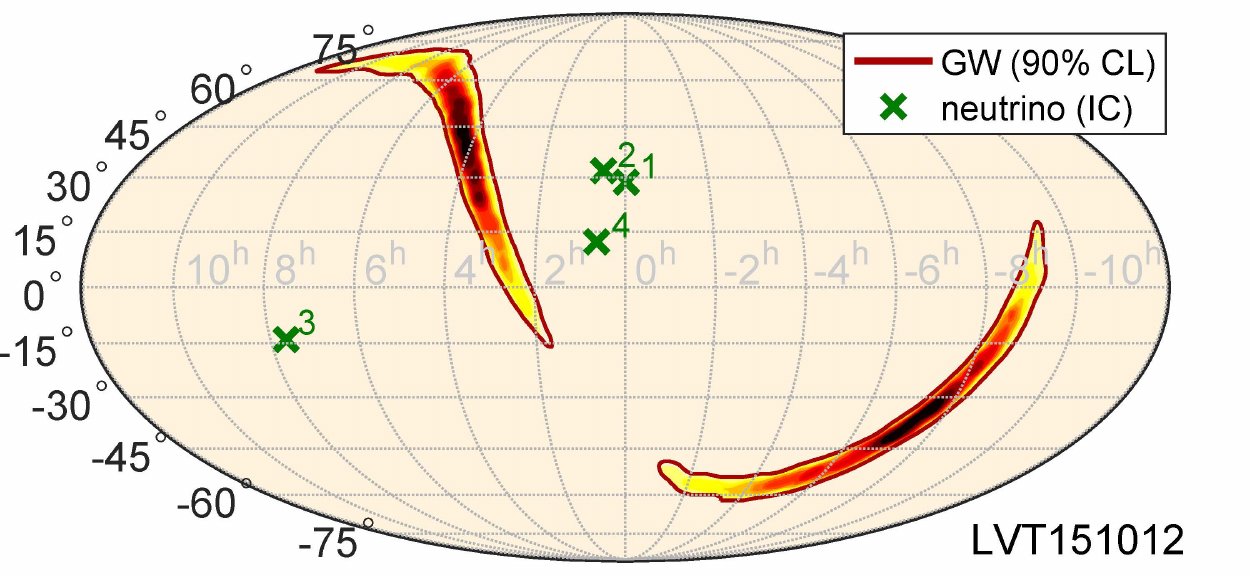}}
\end{center}
\caption{GW skymap for GW151226 (top) and LVT151012 (bottom), and the reconstructed directions for high-energy neutrino candidates detected by IceCube (green crosses) and \textsc{Antares} (blue cross) within $\pm 500$\,s around the GW signals. The maps are in equatorial coordinates. The GW skymap shows the reconstructed probability density contours of the GW event 90\% CL. GW shading indicates the reconstructed probability density of the GW event, darker regions corresponding to higher probability. The neutrino directional uncertainties are below $1^\circ$ for most of the candidates, and in any case too small to be shown.  Neutrino event numbers refer to the first column of Table~\ref{table:neutrinos}.}
\label{fig:skymap}
\end{figure}

We found that of the temporally coincident neutrino candidates, none were directionally coincident with the GW signals at the 90\% credible level, as shown in Fig. \ref{fig:skymap}.

We use the non-detection of joint GW and neutrino events to constrain neutrino emission from the GW source. Since the sensitivity of neutrino detectors is highly dependent on source direction, we calculate upper limits as a function of source direction for the whole sky. 

Upper limits on the neutrino emission for IceCube from a point source within the $\pm500$\,s second interval are calculated in a similar way to the procedure in \cite{2015ApJ...807...46A}, using Monte Carlo simulation to determine the mean fluence required to produce a neutrino signal in 90\% of simulated trials that is above the observed one in the data. For \textsc{Antares}, we computed upper limits (90\% confidence level) using a full Monte Carlo simulation, with the standard \textsc{Antares} chain \cite{brunner100antares,2013NIMPA.725...98M,2016EPJWC.11602002F}, of the detector's response at the time for the GW signal.

\begin{figure}
\begin{center}
\resizebox{0.49\textwidth}{!}{\includegraphics{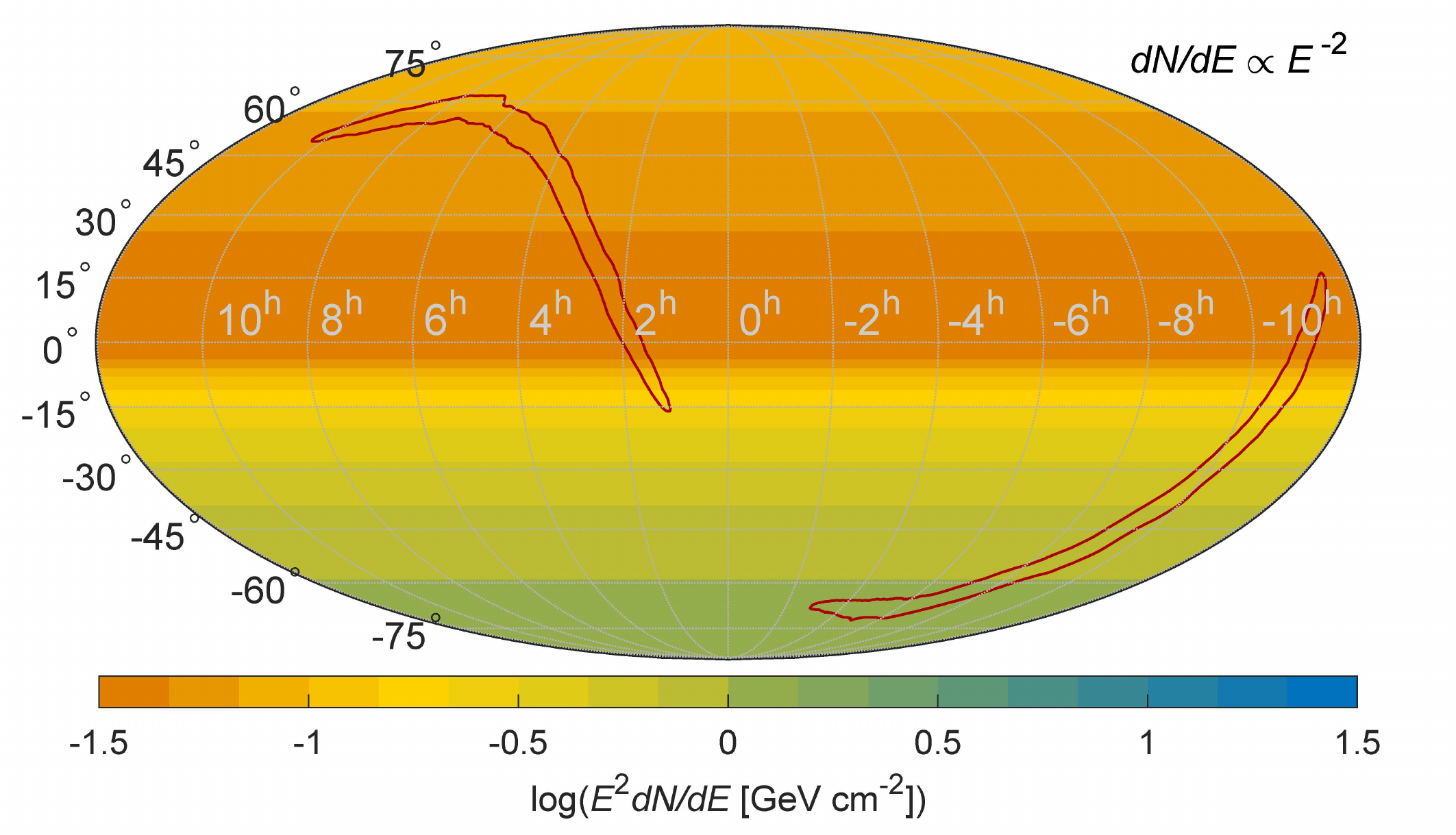}}
\resizebox{0.49\textwidth}{!}{\includegraphics{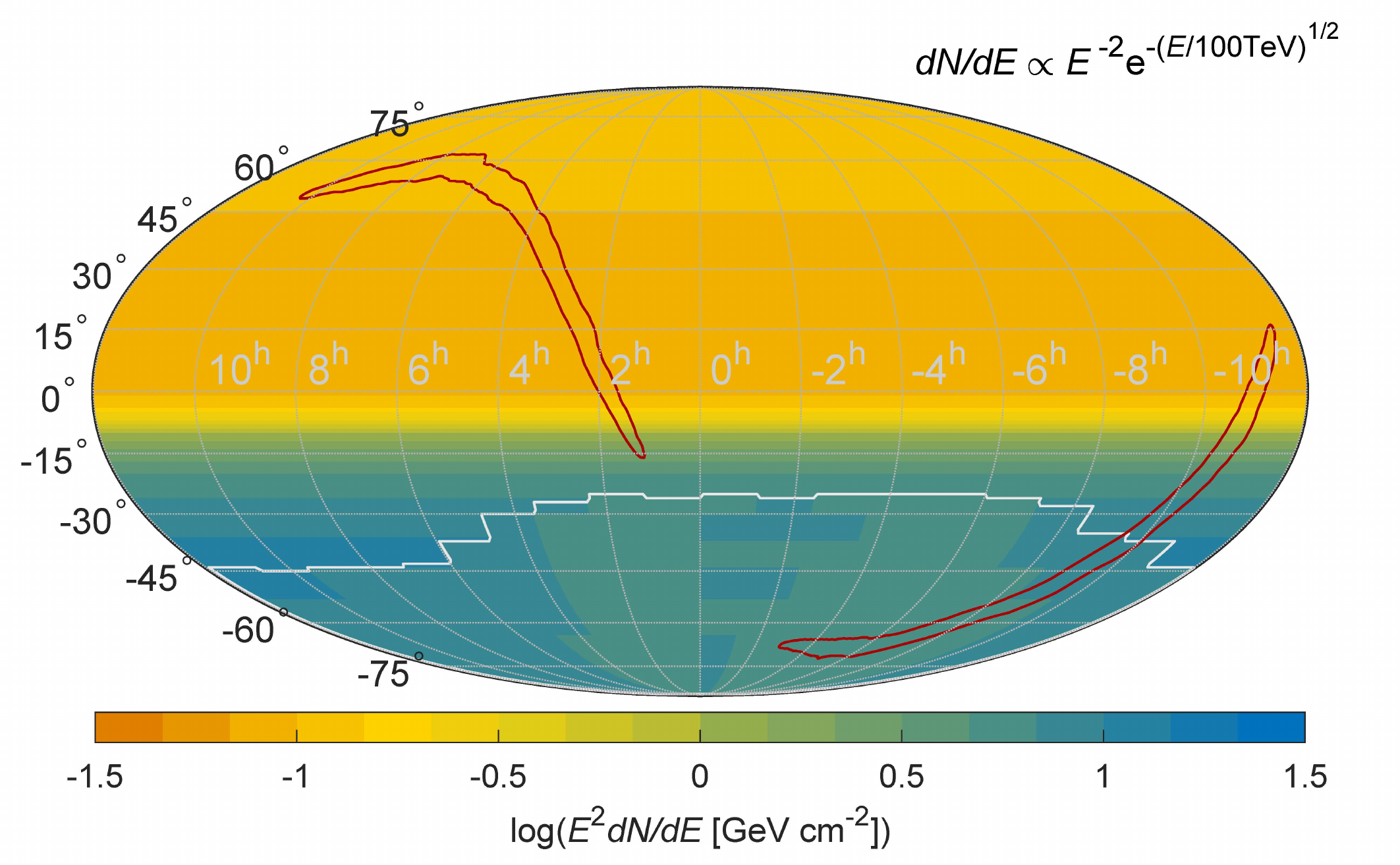}}
\end{center}
\caption{Upper limit for high-energy neutrino spectral fluence ($\nu_{\mu} + \overline{\nu}_{\mu}$) as a function of source direction corresponding to GW151226, assuming $dN/dE\propto E^{-2}$ (top) and $dN/dE\propto E^{-2}\exp[-(E/100\text{TeV})^{1/2}]$ (bottom) neutrino spectra. The region surrounded by a white line shows the part of the sky in which \textsc{Antares} is more sensitive (lowest declinations), while on the rest of the sky, IceCube is more sensitive. For comparison, the 90\% credible-level contour for the GW skymap is also shown.}
\label{fig:UL}
\end{figure}

\begin{figure}
\begin{center}
\resizebox{0.49\textwidth}{!}{\includegraphics{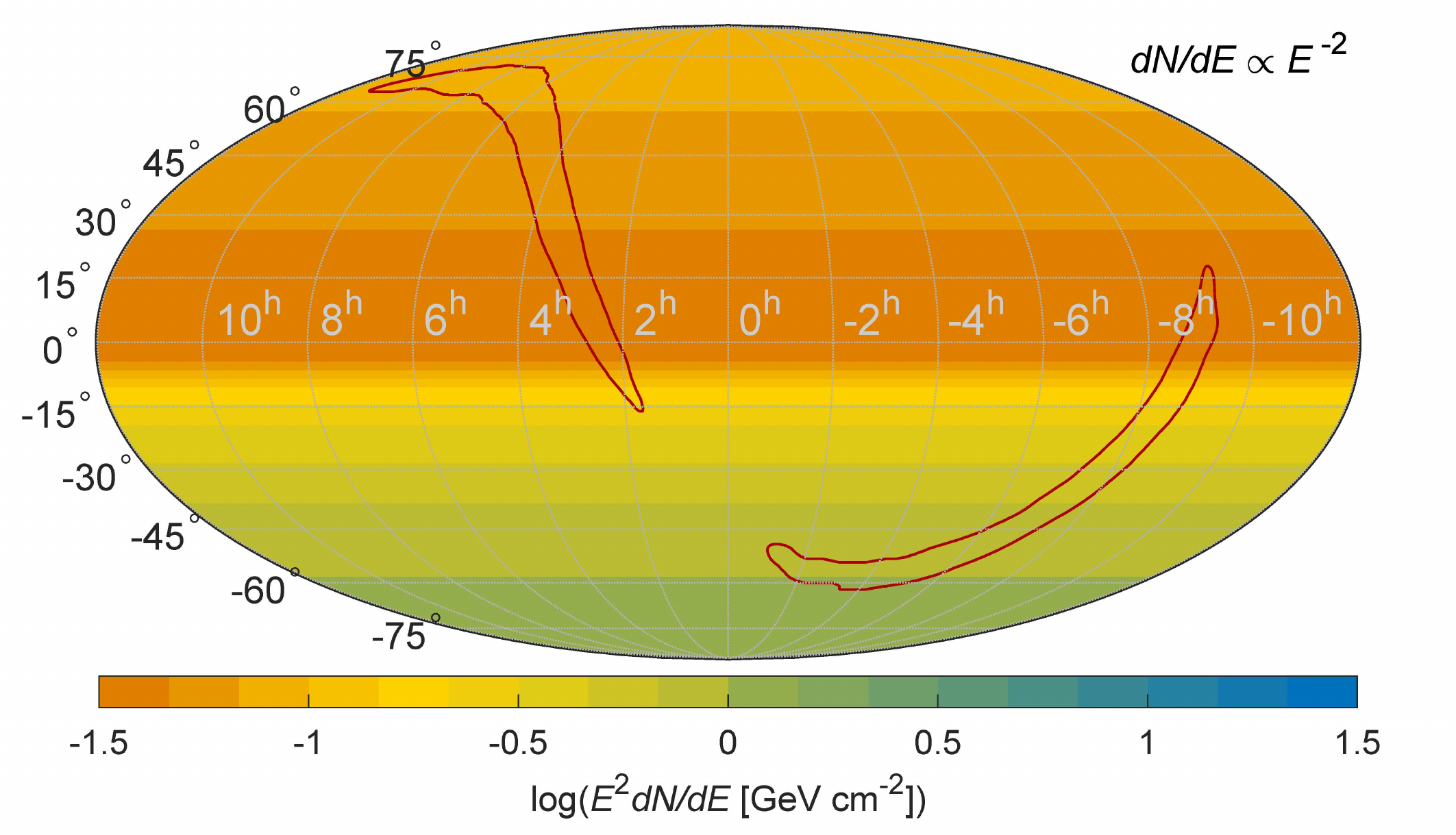}}
\resizebox{0.49\textwidth}{!}{\includegraphics{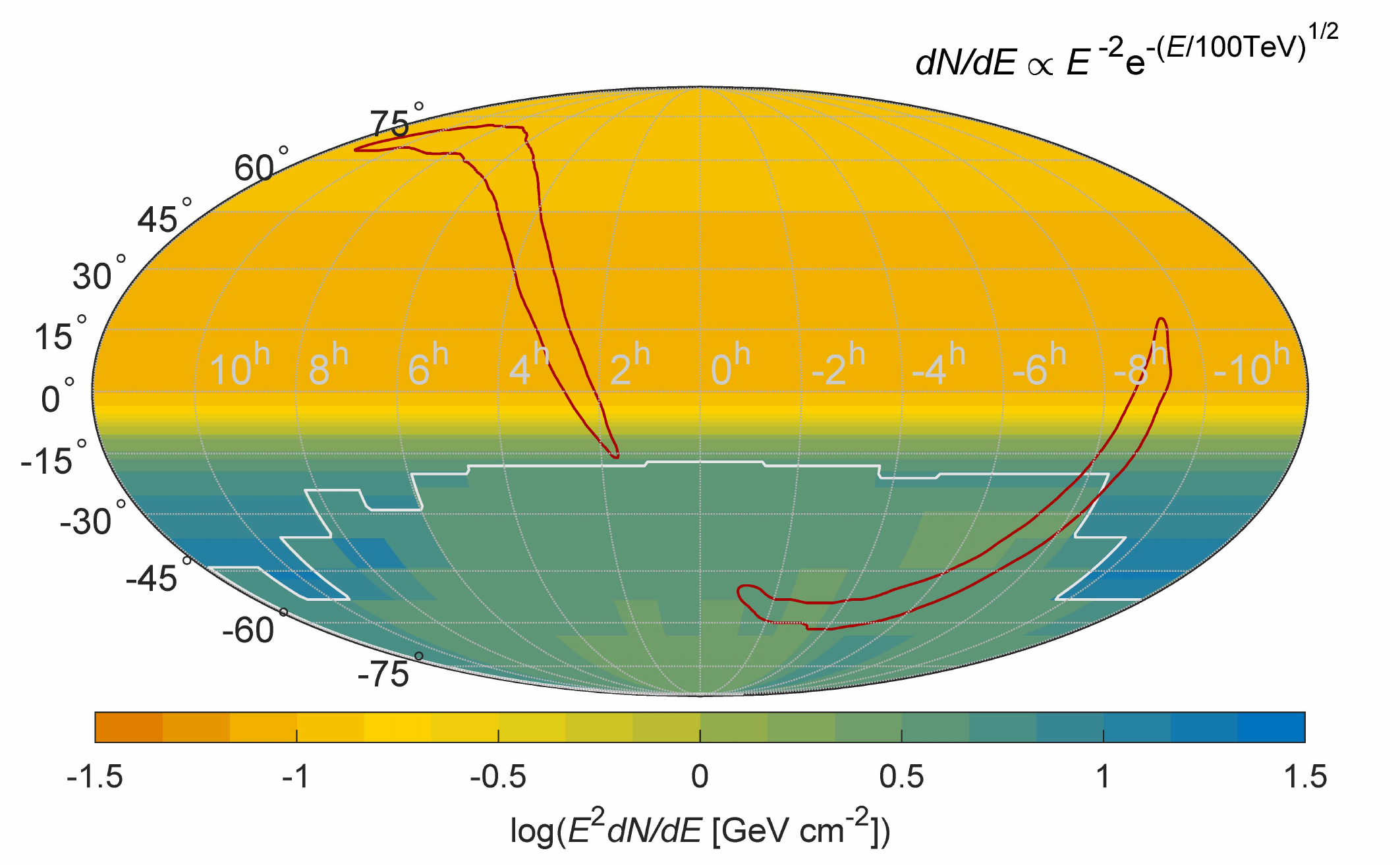}}
\end{center}
\caption{Same as Fig. \ref{fig:UL}, but for LVT151012.}
\label{fig:UL_LVT}
\end{figure}

For a given direction, we adopt the upper limit from IceCube or \textsc{Antares}, whichever is more constraining. Fig. \ref{fig:UL} shows this neutrino spectral fluence upper limit for GW151226 as a function of source direction. We calculate upper limits on the spectral fluence $\phi_0$, for two different neutrino spectral models: $dN/dE=\phi_0 E^{-2}$ typically expected for Fermi acceleration \cite{1997PhRvL..78.2292W}, and $dN/dE=\phi_0 E^{-2}\exp[-(E/100\text{TeV})^{1/2}]$, in order to characterize sensitivity to a source that emits only at lower energies (e.g., \cite{2012PhRvD..86h3007B}). We show the same upper limits for LVT151012 in Fig. \ref{fig:UL_LVT}. These limits are similar to those obtained for GW event GW150914 \cite{2016PhRvD..93l2010A}.

\subsection{Constraints from 3D gravitational wave localization}

The GW signal of a binary merger contains information not only on the source direction, but also on its distance, which can be reconstructed \cite{2016PhRvL.116x1102A}. The source position can therefore be constrained to within a 3D volume \cite{2013ApJ...767..124N}. The GW detectors' direction-dependent sensitivity and the detector noise make such a 3D sky volume skewed towards some directions. Reconstructing a 3D source constraint is useful for identifying possible host galaxies for follow-up observations \cite{2015ApJ...801L...1B,2016ApJ...816...61B,2014ApJ...784....8H,2014ApJ...795...43F,2015ApJS..217....8B}. It can also be used for deriving direction-dependent multimessenger source constraints.

\begin{figure}
\begin{center}
\resizebox{0.49\textwidth}{!}{\includegraphics{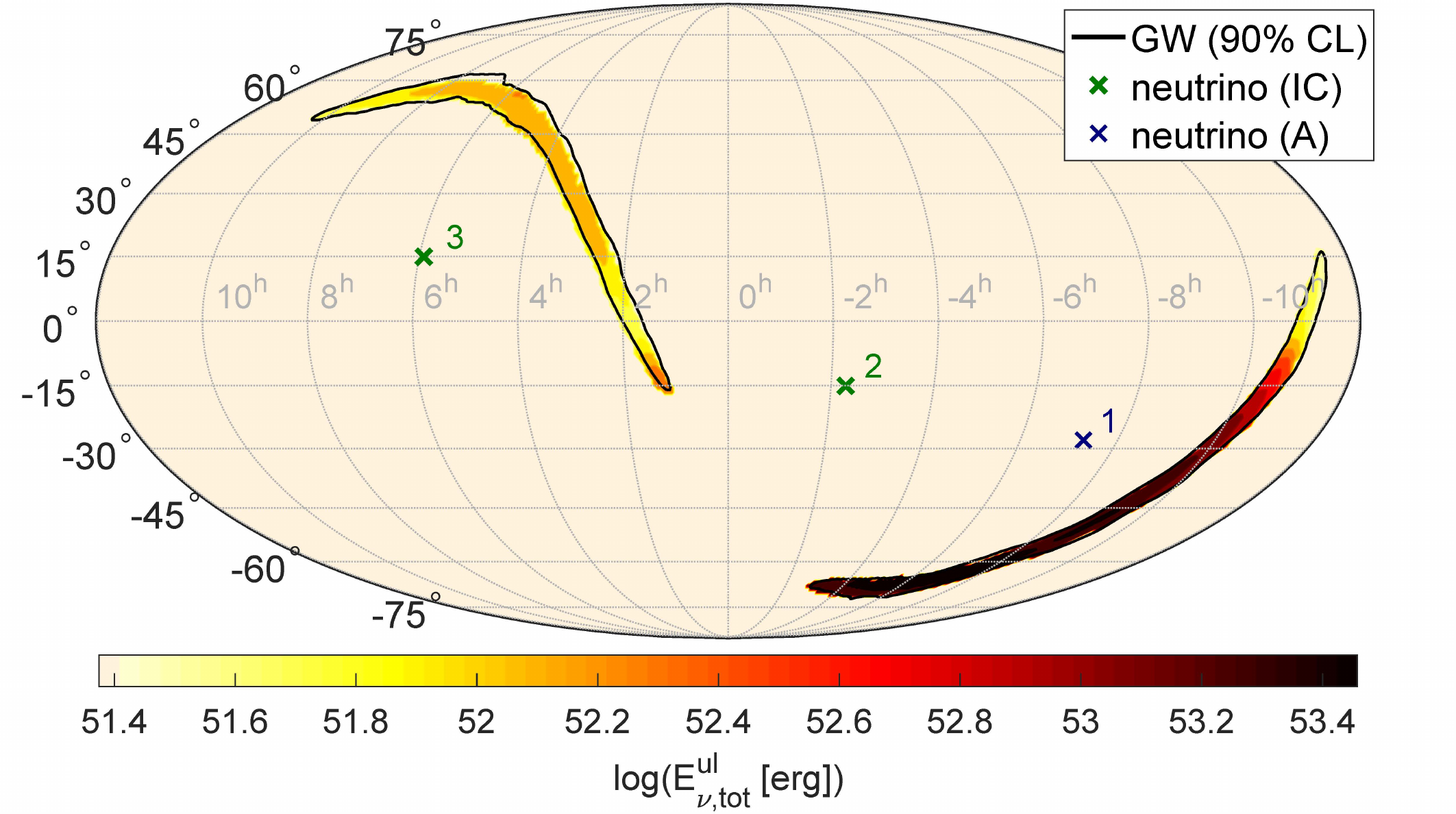}}
\resizebox{0.49\textwidth}{!}{\includegraphics{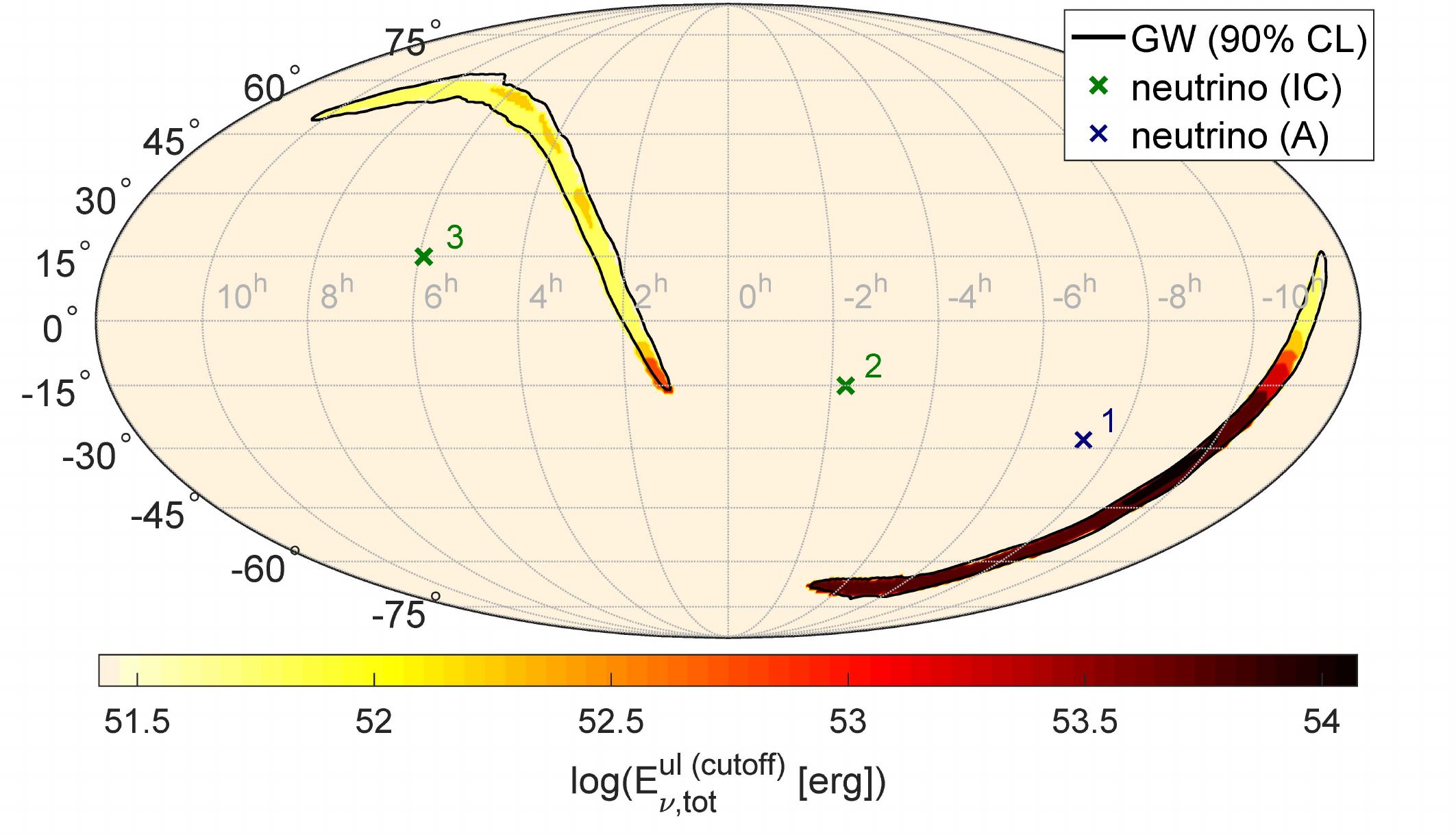}}
\end{center}
\caption{Upper limit on the total energy radiated in high-energy neutrinos by the progenitor of GW151226 as a function of source direction, assuming $dN/dE\propto E^{-2}$ (top) and $dN/dE\propto E^{-2}\exp[-(E/100\text{TeV})^{1/2}]$ (bottom) neutrino spectra. The direction dependent constraint is derived from the direction dependent neutrino spectral fluence upper limit (see Fig. \ref{fig:UL}) as well as the reconstructed 3D GW localization.}
\label{fig:3DUL}
\end{figure}

We adopt the reconstructed sky volume for GW151226 to constrain neutrino emission as a function of source direction \cite{2016ApJ...829L..15S}. We take the lower limit $D_{\rm low}^{95\%}(\vec{x})$ on the source distance for a given direction $\vec{x}$ such that the source is located within this distance at 95\% credible level. We then use $D_{\rm low}^{95\%}(\vec{x})$ to calculate the upper limit on the total isotropic-equivalent energy emitted in neutrinos by the source:
\begin{equation}
\mathrm{E}_{\rm\nu,iso}^{\rm ul}(\vec{x}) = 4\pi \left[D_{\rm low}^{95\%}(\vec{x})\right]^2 \int\frac{dN}{dE} E  dE.
\end{equation}
We obtain upper limits for both $dN/dE\propto E^{-2}$ and $dN/dE\propto E^{-2}\exp[-(E/100\text{TeV})^{1/2}]$ neutrino spectral models. We integrate the spectrum over the interval $[100\,\mbox{GeV}, 100\,\mbox{PeV}]$ for both spectral models. The resulting limits as a function of the position on the sky are shown in Fig. \ref{fig:3DUL}. We see that for either spectral model, there is over two orders of magnitude variation in the total neutrino emission upper limit over the GW skymap. To quantify the range of the upper limits over the skymap, we calculate the minimum and maximum upper limit values over the whole GW skymap, separately for the two spectral models. We obtain the following ranges:
\begin{eqnarray}
\mathrm{E}_{\rm\nu,iso}^{\rm ul} &=& \left(2\times 10^{51} \,\text{--}\, 3 \times 10^{53}\right)\,\text{erg}; \\
\mathrm{E}_{\rm\nu,iso}^{\rm ul \text{(cutoff)}} &=& \left(3\times 10^{51} \,\text{--}\, 2 \times 10^{54}\right)\,\text{erg}.
\end{eqnarray}
For comparison, the total energy emitted from GW151226 in GWs is $\approx 1.8\times 10^{54}$\,erg. Constraints for LVT151012 are about a factor of 4 weaker as its expected distance is about twice that of GW151226 \cite{2016arXiv160604856T}, while both their skymaps similarly lie over a large declination range, corresponding to similar neutrino detector sensitivities.

\section{Conclusion}
\label{sec:conclusion}

Searching in data recorded by the IceCube Neutrino Observatory and the \textsc{Antares} Neutrino Telescope, we detected no neutrino emission associated with the second binary black hole merger, GW151226, discovered by Advanced LIGO. We similarly found no coincident neutrino emission for GW event candidate LVT151012. We used the non-detection to constrain the total neutrino emission from GW151226 to $\sim2\times 10^{51}\text{--}2 \times 10^{54}$\,erg, allowing for different possible neutrino spectra. For these constraints we also adopted 3D GW localizations, and found significant directional dependence in the neutrino emission upper limit. This is due to the fact that the sensitivity of both neutrino and GW detectors is direction dependent.

The observational constraints on total neutrino emission for GW151226 presented here are overall about a factor of two better than the range $5.4\times10^{51}\text{--}3.7\times10^{54}$\,erg previously reported for GW event GW150914 \cite{2016PhRvD..93l2010A}; however, this previous work has not incorporated 3D localization for GWs. Without this change, the range of observational constraints for GW150914 and GW151226 would be essentially identical, since (i) the sensitivities of the neutrino observatories are very similar for the two cases, (ii) the luminosity distance of the two GW events is also similar, and (iii) both GW events have sky localizations consistent with both a northern and southern origin, for which neutrino sensitivities are very different. Nevertheless, the source direction for GW151226 has higher probability of originating from the northern hemisphere, for which the upper limits are significantly more constraining.

High-energy neutrino emission induced by a binary black hole system would require significant gas accretion, as well as for an energetic outflow driven by the accretion disk to be beamed towards the Earth. These conditions are not satisfied for most binary black hole mergers. Nevertheless, with the expected high rate of observations by the Advanced LIGO-Virgo network, neutrino searches can probe even small sub-populations of mergers, testing binary evolution channels in gaseous environments. With the all-sky sensitivity of neutrino detectors, these searches represent a promising way in comprehensively probing high-energy emission also for sources outside of the field of view of electromagnetic telescopes, and even for emission prior to the detection of the GW event.

\begin{acknowledgments}
The authors acknowledge the financial support of the funding agencies:
Centre National de la Recherche Scientifique (CNRS), Commissariat \`a
l'\'ener\-gie atomique et aux \'energies alternatives (CEA),
Commission Europ\'eenne (FEDER fund and Marie Curie Program),
Institut Universitaire de France (IUF), IdEx program and UnivEarthS
Labex program at Sorbonne Paris Cit\'e (ANR-10-LABX-0023 and
ANR-11-IDEX-0005-02), Labex OCEVU (ANR-11-LABX-0060) and the
A*MIDEX project (ANR-11-IDEX-0001-02),
R\'egion \^Ile-de-France (DIM-ACAV), R\'egion
Alsace (contrat CPER), R\'egion Provence-Alpes-C\^ote d'Azur,
D\'e\-par\-tement du Var and Ville de La
Seyne-sur-Mer, France;
Bundesministerium f\"ur Bildung und Forschung
(BMBF), Germany;
Istituto Nazionale di Fisica Nucleare (INFN), Italy;
Stichting voor Fundamenteel Onderzoek der Materie (FOM), Nederlandse
organisatie voor Wetenschappelijk Onderzoek (NWO), the Netherlands;
Council of the President of the Russian Federation for young
scientists and leading scientific schools supporting grants, Russia;
National Authority for Scientific Research (ANCS), Romania;
Mi\-nis\-te\-rio de Econom\'{\i}a y Competitividad (MINECO):
Plan Estatal de Investigaci\'{o}n (refs. FPA2015-65150-C3-1-P, -2-P and -3-P, (MINECO/FEDER)), Severo Ochoa Centre of Excellence and MultiDark Consolider (MINECO), and Prometeo and Grisol\'{i}a programs (Generalitat
Valenciana), Spain;
Ministry of Higher Education, Scientific Research and Professional Training, Morocco.
We also acknowledge the technical support of Ifremer, AIM and Foselev Marine
for the sea operation and the CC-IN2P3 for the computing facilities.
We acknowledge the support from the following agencies:
U.S. National Science Foundation-Office of Polar Programs,
U.S. National Science Foundation-Physics Division,
University of Wisconsin Alumni Research Foundation,
the Grid Laboratory Of Wisconsin (GLOW) grid infrastructure at the University of Wisconsin - Madison, the Open Science Grid (OSG) grid infrastructure;
U.S. Department of Energy, and National Energy Research Scientific Computing Center,
the Louisiana Optical Network Initiative (LONI) grid computing resources;
Natural Sciences and Engineering Research Council of Canada,
WestGrid and Compute/Calcul Canada;
Swedish Research Council,
Swedish Polar Research Secretariat,
Swedish National Infrastructure for Computing (SNIC),
and Knut and Alice Wallenberg Foundation, Sweden;
German Ministry for Education and Research (BMBF),
Deutsche Forschungsgemeinschaft (DFG),
Helmholtz Alliance for Astroparticle Physics (HAP),
Research Department of Plasmas with Complex Interactions (Bochum), Germany;
Fund for Scientific Research (FNRS-FWO),
FWO Odysseus programme,
Flanders Institute to encourage scientific and technological research in industry (IWT),
Belgian Federal Science Policy Office (Belspo);
University of Oxford, United Kingdom;
Marsden Fund, New Zealand;
Australian Research Council;
Japan Society for Promotion of Science (JSPS);
the Swiss National Science Foundation (SNSF), Switzerland;
National Research Foundation of Korea (NRF);
Villum Fonden, Danish National Research Foundation (DNRF), Denmark.
The authors gratefully acknowledge the support of the
United States National Science Foundation (NSF) for the
construction and operation of the LIGO Laboratory and
Advanced LIGO as well as the Science and Technology
Facilities Council (STFC) of the United Kingdom,
the Max-Planck-Society (MPS), and the State of
Niedersachsen/Germany for support of the construction of
Advanced LIGO and construction and operation of the
GEO600 detector. Additional support for Advanced LIGO
was provided by the Australian Research Council. The
authors gratefully acknowledge the Italian Istituto
Nazionale di Fisica Nucleare (INFN), the French Centre
National de la Recherche Scientifique (CNRS), and the
Foundation for Fundamental Research on Matter supported
by the Netherlands Organisation for Scientific Research,
for the construction and operation of the Virgo detector
and the creation and support of the European Gravitational
Observatory (EGO) consortium. The authors also gratefully
acknowledge research support from these agencies as well as
by the Council of Scientific and Industrial Research of India,
Department of Science and Technology, India, Science \&
Engineering Research Board, India; Ministry of Human
Resource Development, India; the Spanish Ministerio de
Econom\'ia y Competitividad, the Conselleria d'Economia i
Competitivitat and Conselleria d'Educaci\'o, Cultura i
Universitats of the Govern de les Illes Balears; the
National Science Centre of Poland; the European
Commission; the Royal Society; the Scottish Funding
Council; the Scottish Universities Physics Alliance; the
Hungarian Scientific Research Fund; the Lyon Institute of
Origins; the National Research Foundation of Korea;
Industry Canada and the Province of Ontario through the
Ministry of Economic Development and Innovation; the
Natural Science and Engineering Research Council
Canada; Canadian Institute for Advanced Research; the
Brazilian Ministry of Science, Technology, and
Innovation, Fundação de Amparo \`{a} Pesquisa do Estado de
São Paulo (FAPESP); Russian Foundation for Basic
Research; the Leverhulme Trust; the Research
Corporation; Ministry of Science and Technology,
Taiwan; and the Kavli Foundation. The authors gratefully
acknowledge the support of the NSF, STFC, MPS, INFN,
CNRS, and the State of Niedersachsen/Germany for the
provision of computational resources.
\end{acknowledgments}

\bibliographystyle{h-physrev}

%
  %
  \let\author\myauthor
  \let\affiliation\myaffiliation
  \let\maketitle\mymaketitle
  \title{Authors}
  \pacs{}
   \include{author_list_GWHEN_new_format1}
   \include{IceCube_authorlist}
   \include{LVC_authorlist}
  \newpage
  \maketitle

\end{document}

%% file: author_list_GWHEN_new_format1.tex
\author{A.~Albert}
\affiliation{{GRPHE - Universit\'e de Haute Alsace - Institut universitaire de technologie de Colmar, 34 rue du Grillenbreit BP 50568 - 68008 Colmar, France}}

\author{M.~Andr\'e}
\affiliation{{Technical University of Catalonia, Laboratory of Applied Bioacoustics, Rambla Exposici\'o, 08800 Vilanova i la Geltr\'u, Barcelona, Spain}}

\author{M.~Anghinolfi}
\affiliation{{INFN - Sezione di Genova, Via Dodecaneso 33, 16146 Genova, Italy}}

\author{G.~Anton}
\affiliation{Erlangen Centre for Astroparticle Physics, Friedrich-Alexander-Universit\"at Erlangen--N\"urnberg, Erwin-Rommel-Str.~1, 91058 Erlangen, Germany}

\author{M.~Ardid}
\affiliation{{Institut d'Investigaci\'o per a la Gesti\'o Integrada de les Zones Costaneres (IGIC) - Universitat Polit\`ecnica de Val\`encia. C/  Paranimf 1, 46730 Gandia, Spain}}

\author{J.-J.~Aubert}
\affiliation{{Aix Marseille Univ, CNRS/IN2P3, CPPM, Marseille, France}}

\author{T.~Avgitas}
\affiliation{APC, AstroParticule et Cosmologie, Universit\'e Paris Diderot, CNRS/IN2P3, CEA/Irfu, Observatoire de Paris, Sorbonne Paris Cit\'e, F-75205 Paris Cedex 13, France}

\author{B.~Baret}
\affiliation{APC, AstroParticule et Cosmologie, Universit\'e Paris Diderot, CNRS/IN2P3, CEA/Irfu, Observatoire de Paris, Sorbonne Paris Cit\'e, F-75205 Paris Cedex 13, France}

\author{J.~Barrios-Mart\'{\i}}
\affiliation{{IFIC - Instituto de F\'isica Corpuscular (CSIC - Universitat de Val\`encia) c/ Catedr\'atico Jos\'e Beltr\'an, 2 E-46980 Paterna, Valencia, Spain}}

\author{S.~Basa}
\affiliation{{LAM - Laboratoire d'Astrophysique de Marseille, P\^ole de l'\'Etoile Site de Ch\^ateau-Gombert, rue Fr\'ed\'eric Joliot-Curie 38,  13388 Marseille Cedex 13, France}}

% ...

\collaboration{ANTARES Collaboration}
\noaffiliation 

%% file: IceCube_authorlist.tex
\author{M.~G.~Aartsen}
\affiliation{Department of Physics, University of Adelaide, Adelaide, 5005, Australia}
\author{M.~Ackermann}
\affiliation{DESY, D-15735 Zeuthen, Germany}
\author{J.~Adams}
\affiliation{Dept.~of Physics and Astronomy, University of Canterbury, Private Bag 4800, Christchurch, New Zealand}
\author{J.~A.~Aguilar}
\affiliation{Universit\'e Libre de Bruxelles, Science Faculty CP230, B-1050 Brussels, Belgium}
\author{M.~Ahlers}
\affiliation{Dept.~of Physics and Wisconsin IceCube Particle Astrophysics Center, University of Wisconsin, Madison, WI 53706, USA}
\author{M.~Ahrens}
\affiliation{Oskar Klein Centre and Dept.~of Physics, Stockholm University, SE-10691 Stockholm, Sweden}
\author{I.~Al~Samarai}
\affiliation{D\'epartement de physique nucl\'eaire et corpusculaire, Universit\'e de Gen\`eve, CH-1211 Gen\`eve, Switzerland}
\author{D.~Altmann}
\affiliation{Erlangen Centre for Astroparticle Physics, Friedrich-Alexander-Universit\"at Erlangen--N\"urnberg, Erwin-Rommel-Str.~1, 91058 Erlangen, Germany}
\author{K.~Andeen}
\affiliation{Department of Physics, Marquette University, Milwaukee, WI, 53201, USA}
\author{T.~Anderson}
\affiliation{Dept.~of Physics, Pennsylvania State University, University Park, PA 16802, USA}
% ...

\collaboration{IceCube Collaboration}
\noaffiliation 

%% file: LVC_authorlist.tex
\author{B.~P.~Abbott}
\affiliation{LIGO, California Institute of Technology, Pasadena, CA 91125, USA}

\author{R.~Abbott}
\affiliation{LIGO, California Institute of Technology, Pasadena, CA 91125, USA}

\author{T.~D.~Abbott}
\affiliation{Louisiana State University, Baton Rouge, LA 70803, USA}

\author{M.~R.~Abernathy}
\affiliation{American University, Washington, D.C. 20016, USA}

\author{F.~Acernese}
\affiliation{Universit\`a di Salerno, Fisciano, I-84084 Salerno, Italy}
\affiliation{INFN, Sezione di Napoli, Complesso Universitario di Monte S.Angelo, I-80126 Napoli, Italy}

\author{K.~Ackley}
\affiliation{University of Florida, Gainesville, FL 32611, USA}

\author{C.~Adams}
\affiliation{LIGO Livingston Observatory, Livingston, LA 70754, USA}

\author{T.~Adams}
\affiliation{Laboratoire d'Annecy-le-Vieux de Physique des Particules (LAPP), Universit\'e Savoie Mont Blanc, CNRS/IN2P3, F-74941 Annecy-le-Vieux, France}

\author{P.~Addesso}
\affiliation{University of Sannio at Benevento, I-82100 Benevento, Italy and INFN, Sezione di Napoli, I-80100 Napoli, Italy}

\author{R.~X.~Adhikari}
\affiliation{LIGO, California Institute of Technology, Pasadena, CA 91125, USA}

% ...

\collaboration{LIGO Scientific Collaboration and Virgo Collaboration}
\noaffiliation